\documentclass[prb,twocolumn,showpacs,superscriptaddress,amsmath,amssymb]{revtex4}

\usepackage{graphicx}
\usepackage{latexsym}
\usepackage{amsmath}
\usepackage{amssymb}
\usepackage{amsfonts}
\usepackage{color}

\newcommand{\ds}{\displaystyle}

\begin{document}
\bibliographystyle{apsrev}

\title{Triplet proximity effect in superconducting heterostructures with a half-metallic layer}

\author{S. Mironov}
\affiliation{University Bordeaux, LOMA UMR-CNRS 5798, F-33405 Talence Cedex, France}
\affiliation{Institute for Physics of Microstructures, Russian Academy of Sciences, 603950 Nizhny Novgorod, GSP-105, Russia}
\author{A. Buzdin}
\affiliation{University Bordeaux, LOMA UMR-CNRS 5798, F-33405 Talence Cedex, France}

\date{\today}
\begin{abstract}
We present the Usadel theory describing the superconducting proximity effect in heterostructures with a half-metallic layer. It is shown that the full spin polarization inside the half-metals gives rise to the giant triplet spin-valve effect in superconductor (S) -- ferromagnet (F) -- half-metal (HM) trilayers as well as to the $\varphi_0$-junction formation in the S/F/HM/F/S systems. In addition, we consider the exactly solvable model of the S/F/HM trilayers of atomic thickness and demonstrate that it reproduces the main features of the spin-valve effect found within the Usadel approach. Our results are shown to be in a qualitative agreement with the recent experimental data on the spin-valve effect in ${\rm MoGe/Cu/Ni/CrO_2}$ hybrids [A. Singh {\it et al.}, Phys. Rev. X \textbf{5}, 021019 (2015)].
\end{abstract}

\pacs{74.62.-c, 74.78.Fk, 74.45.+c, 72.25.-b}

\maketitle

\section{Introduction}

Spin-polarized superconducting states attracts growing interest since they are expected to provide powerful mechanisms for controlling the current and magnetization in the devices of superconducting spintronics.\cite{Linder_Review} Although polarized states are not supported in conventional s-wave superconductors, they can emerge in artificial heterostructures consisting of a superconductor (S) and several ferromagnetic (F) layers with different orientations of magnetic moments. \cite{Bergeret_2001,Kadigrobov,Buzdin_RMP,Bergeret_RMP} The non-collinear exchange field in the F-layers destroys the spin-singlet structure of Cooper pairs penetrating from the superconductor. This results in the appearance of spin-triplet superconducting correlations with all possible spin projections $S$, both polarized states with $S=\pm 1$ and non-polarized one with $S=0$.

The correlations with $S\pm 1$ have two distinctive features which give an insight into their experimental observation and practical utilization.\cite{Bergeret_2001,Kadigrobov} First, such correlations are unsensitive to the exchange field parallel to the spin quantization axis. As a result, they become long-range: in diffusive systems their decay length inside the ferromagnet is comparable with the one in normal metal while the correlations with $S=0$ decay at much shorter distances from the superconductor. Second, these long-range triplet correlations (LRTC) appear only if the ferromagnet has a non-collinear distribution of magnetization. Thus, to control the amplitude of the LRTC in the system one can use ferromagnetic bilayer with tunable mutual orientation of magnetic moments while the effects coming from the correlations with $S=0$ can be damped by increasing the F-layer thickness.

Experimental observation of the LRTC is mostly based on probing the long-range Josephson current in the ${\rm S/F^\prime/F/F^\prime/S}$ junctions.\cite{Khaire, Sosnin, Robinson2, Sprungmann, Wang, Khasawneh} If the thickness of the central F layer is believed to exceed the decay length of the short-range non-polarized correlations the observation of non-zero critical current can be attributed to the presence of the LRTC.\cite{Houzet_Buzdin} Other observations of the LRTC are based on the so-called triplet spin-valve effect in ${\rm S/F_1/F_2}$ and ${\rm F_1/S/F_2}$ systems revealing in non-monotonic dependence of the S-layer critical temperature $T_c$ on the angle $\theta$ between the magnetic moments in the ${\rm F_1}$ and ${\rm F_2}$ layers.\cite{Leksin,Zdravkov,Jara,Flokstra,Dybko} The LRTC open an additional channel for the ``leakage" of the Cooper pairs from the superconductor. As a result, $T_c$ can have the minimum at $\theta\not= 0,\pi$.\cite{Fominov_SFF,Mironov_FSF} However this effect in $T_c(\theta)$ is typically washed out by the monotonically increasing contribution from the correlations with $S=0$, which shifts the minimum of $T_c$ from $\theta=\pi/2$.

During the past few years the focus in the studies of LRTC is moving towards the heterostructures containing half-metallic (HM) layers (e.g. ${\rm CrO_2}$).\cite{Pickett,Coey} A recent progress in fabrication of such structures has resulted in several breakthrough experiments manifesting the long-range Josephson current through the layer of ${\rm CrO_2}$\cite{Keizer,Anwar} and triplet spin-valve effect in ${\rm MoGe/Cu/Ni/CrO_2}$ structures.\cite{Aarts} The importance of these experiments is connected with the fact that in half-metals the energy bands for electrons with spin up and down are separated at a distances comparable with the Fermi energy. As a result, only spin-polarized correlations with $S=+1$ can penetrate into the HM-layer while all other correlations should vanish at its boundary and cannot influence the Josephson current in ${\rm S/F/HM/F/S}$ junctions or the triplet spin-valve effect in S/F/HM systems. Thus, half-metals provide a unique possibility to probe the phenomena caused by LRTC independently from other effects.

However up to now there is no convenient and commonly accepted theoretical model describing the superconducting proximity effect with half-metals. Most common approaches for the treatment of the proximity effect in multilayered structures are based on the quasiclassical approximation.\cite{Buzdin_RMP,Bergeret_RMP} This approximation becomes broken near the interfaces which implies using some sort of boundary conditions matching the quasiclassical Green functions in different layers. In contrast with the S/F interfaces where near the critical temperature one may use linear Kupriyanov-Lukichev boundary conditions for the anomalous Green function,\cite{Kupriyanov} the interfaces with half-metals require more sophisticated boundary conditions since the number of the Green function components at the opposite sides of the interface is different due to the large energy separation of the spin-up and spin-down bands. There are a lot of papers where the authors made attempts to overcome this problem using different versions of the scattering matrix approach. The resulting boundary conditions were extensively used for the description of half-metals within the Blonder-Tinkham-Klapwijk,\cite{Zheng,Feng,Linder_HM,Enoksen} Bogoliubov-de-Gennes,\cite{Asano_HM,Sawa,Beri} Eilenberger,\cite{Tokuyasu,Eschrig_2003,Eschrig_2008,Eschrig_2009,Eschrig_2010} and Usadel\cite{Cottet_U,Bergeret_HM_R,Bergeret_HM,Eschrig2015} formalisms. Despite the fact that these models provide a number of generic qualitative predictions, most of the results strongly depend on the microscopical mechanisms of the singlet-triplet conversion and on the concrete form of the scattering matrices,\cite{Eschrig_2008,Takahashi,Kupferschmidt} which are not available form the experimental data. An alternative phenomenological approach based on the circuit theory\cite{Nazarov} does not contain any information about the particular geometry of the system and, thus, can hardly be applied for the quantitative description of real heterostructures.

In the present paper we propose the phenomenological model of the superconducting proximity effect with half-metals based on the Usadel equation in the diffusive limit. Our model is based on the following three key assumptions. (i) The impurities are non-magnetic and do not cause spin-flip processes, which allows to introduce the anomalous Green function inside the half-metal. The Green function component with $S=+1$ satisfies the Usadel equation equivalent to the one in normal metal while all other components are zero. (ii) The are no barriers at the boundaries of the HM layers which results in the continuity of the component with $S=+1$. (iii) The components with $S=0$ and $S=-1$ cannot penetrate the HM layer and vanish at its outer boundaries. Our simple model is shown to explain all main features of the recently observed triplet spin-valve effect in ${\rm MoGe/Cu/Ni/CrO_2}$ structures \cite{Aarts} and predict several unusual phenomena manifesting the differences between the influence of weak and strong ferromagnets on the proximity effect.

To verify our main conclusions we also considered the situation when the layers of the S/F/HM spin-valve have atomic thickness and are separated by the tunnel barriers. The advantage of such microscopical model is the possibility to find the exact solution of the Gor'kov equations without any prior assumptions about the profiles of the Green functions. Using this model we obtained the analytical dependencies $T_c(\theta)$ which appear to reproduce all main features found within the Usadel approach.

The paper is organized as follows. In Sec.~\ref{Sec_Model} we introduce our model. In Sec.~\ref{Sec_SFHM} we apply it for the description of the triplet spin-valve effect in S/F/HM systems and compare our results with the experimental observations of Ref.~\onlinecite{Aarts}. In Sec.~\ref{Sec_SFHMFS} we study the anomalous Josephson effect in dirty S/F/HM/F/S structures with non-coplanar magnetic moments and demonstrate that such systems support the states with the spontaneous phase difference, previously predicted for the ballistic limit.\cite{Eschrig_2009} In Sec.~\ref{Sec_Atomic} we consider the spin-valve effect for the S/F/HM systems of atomic thickness and compare the results with the conclusions of Sec.~\ref{Sec_SFHM}. In Sec.~\ref{Sec_Conc} we summarize our results and discuss their possible applications.

\section{Model}\label{Sec_Model}

\begin{figure}[t!]
\includegraphics[width=0.25\textwidth]{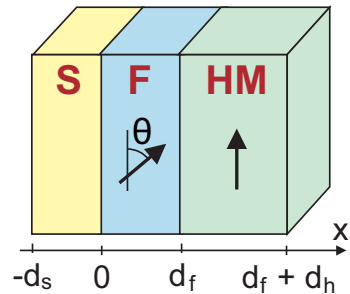}
\caption{(Color online) The sketch of the ${\rm S/F/HM}$ spin valve. The exchange field in the F-layer makes the angle $\theta$ with the spin quantization axis in the half-metal.} \label{Fig_SFHM}
\end{figure}

Let us consider a multilayered structure consisting of superconductors, ferromagnets and half-metals with the interfaces perpendicular to the $x$-axis. Two examples of such structures are shown in Fig.~\ref{Fig_SFHM} and Fig.~\ref{Fig_SFHMFS}. We assume that the systems is in the diffusive limit and the temperature $T$ is close to the critical temperature $T_c$ of the superconducting transition. In this case outside the half-metallic layers the superconducting properties of the system can be described in terms of the linearized Usadel equation\cite{Eschrig}
\begin{equation}  \label{Main_System}
\frac{D}{2}\partial_x^2\hat{f}-\omega_n \hat{f}-\frac{i}{2}\left(\mathbf{h}
\mathbf{\hat{\sigma}} \hat{f}+ \hat{f}\mathbf{h}\mathbf{\hat{\sigma}}\right)+
\hat{\Delta}=0,
\end{equation}
where the quasiclassical Green function
\begin{equation}  \label{Matrix_f}
\hat{f}=\left(
                \begin{array}{cc}
                  f_{\uparrow\uparrow} & f_{\uparrow\downarrow} \\
                  f_{\downarrow\uparrow} & f_{\downarrow\downarrow} \\
                \end{array}
              \right)
=\left(f_s+\mathbf{f}_t\hat{\sigma}\right)i\hat{\sigma}_y
\end{equation}
is the $2\times 2$ matrix in the spin space, $\hat{\Delta}=\Delta i\hat{\sigma}_y$ is the superconducting pairing potential inside the S layers, $\omega_n=\pi T(2n+1)>0$ are the Matsubara frequencies, ${\bf h}$ is the exchange field in the ferromagnets, and $D$ is the diffusion constant.

To describe the superconducting correlations inside the half-metallic layers we assume that (i) electron scattering on the impurities does not cause spin-flips and (ii) the barriers are not spin-active. In this case we may introduce the Usadel Green function with only one nonzero component $f_{\uparrow\uparrow}$ meaning that the spin polarization in the half-metal is directed along the $z$-axis.

%%%%%%%%%%%%%%%%%%%%%%%%%%%%%%%%%%%%%%%%%%%%
%%%%%%%%%%%%%%%%%%%%%%%%%%%%%%%%%%%%%%%%%%%%

To match the solutions of the Eq.~(\ref{Main_System}) in different layers one should put the boundary conditions at each interface. We assume all interfaces between the layers to be transparent for electrons. In this case at the interfaces which separate two non-halfmetallic material the boundary conditions come to the continuity of $\hat{f}$ and the combination $\sigma\partial_x\hat{f}$ ($\sigma$ is the Drude conductivity of the corresponding layer).\cite{Kupriyanov} Similarly, for the interfaces with the half-metal we demand only the continuity of $f_{\uparrow\uparrow}$ and $\sigma\partial_x f_{\uparrow\uparrow}$ while all other components $f_{\downarrow\downarrow}$, $f_{\uparrow\downarrow}$ and $f_{\downarrow\uparrow}$ should vanish at the HM-layer boundaries. At the outer boundaries of the heterostructure we demand $\partial_x\hat{f}=0$.

The physical meaning of the boundary conditions at the interface with half-metal is very clear: the interface with the HM layer plays the role of spin filter, which is absolutely transparent for the correlations with $S=+1$ and opaque for all other correlations.

\section{Triplet spin-valve effect in S/F/HM systems}\label{Sec_SFHM}

The geometry of the S/F/HM system under consideration is shown schematically in Fig.~\ref{Fig_SFHM}. We choose the origin of the $x$-axis in a way that the superconductor is at $-d_s<x<0$, the ferromagnet is at $0<x<d_f$, and the HM layer occupies the region $d_f<x<d_f+d_h$. The exchange field ${\bf h}$ in the F-layer is assumed to be rotated on the angle $\theta$ in the $xz$-plane and have two components: $h_{z}=h \cos\theta$ and $h_{x}=h \sin\theta$. The spin quantization axis in the HM-layer coincides with the $z$ one.

To demonstrate the key difference between the S/F/HM system and classical ${\rm S/F_2/F_2}$ spin valves it is instructive to write the explicit form of Eq.~(\ref{Main_System}) inside the F-layer for all components $f_s$ and ${\bf f}_t$:
\begin{equation}  \label{Usadel_F}
\begin{array}{l}{\ds (D_f/2)\partial_x^2 f_s=\omega_n f_s+ih\cos\theta f_{tz}+ih\sin\theta f_{tx},
}\\{}\\{\ds (D_f/2)\partial_x^2 f_{tz}=\omega_n f_{tz}+ih\cos\theta f_s,}\\{}\\{\ds (D_f/2) \partial_x^2 f_{tx}=\omega_n f_{tx}+ih\sin\theta f_s,}\\{}\\{\ds (D_f/2)\partial_x^2 f_{ty}=\omega_n f_{ty},} \end{array}
\end{equation}
where $D_f$ is the diffusion constants in the ferromagnet. Clearly the spin-singlet component $f_s$ of the anomalous function coming from the superconductor induces the triplet components $f_{tz}$ and $f_{tx}$ while the equation for $f_{ty}$ remains independent. In the ${\rm S/F_2/F_2}$ structures the boundary conditions do not mix different components of the function $\hat{f}$ and, thus, the absence of the source in equation for $f_{ty}$ immediately leads to $f_{ty}(x)= 0$ in the whole heterostructure.

However, in the S/F/HM systems the situation is completely different. The penetration of the non-zero component $f_{\uparrow\uparrow}=-f_{tx}+if_{ty}$ into the HM-layer together with the vanishing of the component $f_{\downarrow\downarrow}=f_{tx}+if_{ty}$ is possible only if inside the half-metal
\begin{equation}\label{Fxy}
f_{ty}=if_{tx}\not= 0.
\end{equation}
Consequently, this results in the appearance of $f_{ty}$ also in the S and F layers.

For the further analysis it is convenient to exclude $f_{ty}$ from (\ref{Main_System}) by substituting the solution of the equation for $f_{ty}$ into the boundary conditions. As a result, all information about the component $f_{ty}$ becomes included into the effective boundary condition for $f_{tx}$, which reads (see Appendix \ref{App_BoundCond})
\begin{equation}\label{fx_bound}
\left.\frac{\partial_x f_{tx}}{f_{tx}}\right|_{x=d_f}=-q_f\Gamma
\end{equation}
with
\begin{equation}\label{Gamma_def}
\Gamma=2\mu_h+\frac{\mu_s+\mu_f}{1+\mu_s \mu_f}.
\end{equation}
In this expression
\begin{equation}\label{mu_def}
\mu_j=\frac{\sigma_j}{\sigma_f}\sqrt{\frac{D_f}{D_j}}\tanh(q_jd_j),
\end{equation}
where the index $j\in\left\{s,f,h\right\}$ corresponds to the S, F and HM layers respectively, $q_j=\sqrt{2\omega_n/D_j}$, $D_j$ and $\sigma_j$ are the diffusion constant and the normal conductivity in the $j$-th layer.

Note that in the case when the diffusion constants and conductivities of all layers are equal to each other the expression for $\Gamma$ takes the form
\begin{equation}\label{fx_bound_ident}
\Gamma=2\tanh(q_hd_h)+\tanh\left[q_f(d_s+d_f)\right].
\end{equation}
Clearly, this expression reflects the fact that at the F/HM interface the component $f_{tx}$ of the Green function induces two components ($f_{tx}$ and $f_{ty}$) in the half-metal and also the component $f_{ty}$ in the S/F bilayer.

Now let us analyze the dependence of the S-layer critical temperature on the angle $\theta$. To simplify the calculations we assume that the thickness of the S-layer is much smaller than the superconducting coherence length $\xi_{s0}=\sqrt{D_s/4\pi T_{c0}}$, which enables neglecting the spatial variation of the pairing potential $\Delta$ across the superconducting film (here $T_{c0}$ is the critical temperature of the isolated superconductor). For convenience we choose $\Delta$ to be real. Then the dependence $T_c(\theta)$ is defined by the self-consistency equation
\begin{equation}\label{SelfConsistencyEquation}
\Delta\ln\frac{T_c\left(\theta\right)}{T_{c0}}+\sum\limits_{n=0}^{\infty}
\left[\frac{\Delta}{n+1/2}-2\pi T_c\left(\theta\right) f_s\left(\theta\right)\right]=0.
\end{equation}

The equation (\ref{Main_System}) with the boundary conditions described above allows us to obtain the analytical expressions for the anomalous Green function in each layer (the details of the calculations are presented in Appendix \ref{App_Calc}). Inside the superconductor the component $f_s$ has the form
\begin{equation}\label{fs_res}
f_s=\Re\left(\frac{\Delta}{\omega_n+\tau_{\pi}^{-1}}\right)- \frac{\Delta}{\omega_n}  \frac{\mu_sW^2\sin^2\theta}{Q\sin^2\theta +\Gamma+\left(\Gamma-2\mu_h\right)\cos^2\theta},
\end{equation}
where $q=\sqrt{2\left(\omega_n+ih\right)/D_f}$, $p=\mu_sq_f$,
\begin{equation}\label{W_def_main}
W=\Im\left\{\frac{q}{q\cosh(qd_f)+p\sinh(qd_f)}\right\},
\end{equation}
\begin{equation}\label{Q_def_main}
Q=\Re \left\{\frac{q}{q_f}\frac{p+q\tanh(qd_f)}{q+p\tanh(qd_f)}\right\},
\end{equation}
and the pair-breaking parameter
\begin{equation}\label{tau_pi_def}
\tau_\pi^{-1}=\frac{\sigma_f}{\sigma_s}\frac{D_s}{2d_s}q\coth(qd_f)
\end{equation}
is the same as for the S/F/S junction with the F-layer thickness $2d_f$ in the $\pi$-state (see, e.g., Ref.~\onlinecite{Buzdin_RMP}).

%%%%%%%%%%%%%%%%%%%%%%%%%%%%%%%%%%%%%%%%%%%%
\begin{figure}[t!]
\includegraphics[width=0.38\textwidth]{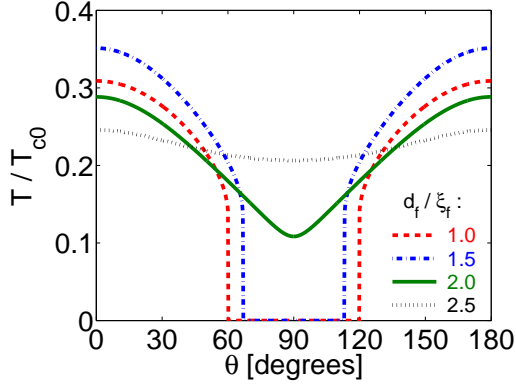}
\caption{(Color online) The dependencies of the critical temperature on the angle $\theta$ for different thicknesses $d_f$ of the ferromagnetic layer. The system parameters are: $d_s=0.5\xi_{s0}$, $h/4\pi T_{c0}=2$, $d_h=\sqrt{D_h/4\pi T_{c0}}$, $D_f/D_s=\sigma_f/\sigma_s=10^{-3}$, and $D_h/D_f=\sigma_h/\sigma_f=10^{-3}$.} \label{Fig_Tcphi}
\end{figure}
%%%%%%%%%%%%%%%%%%%%%%%%%%%%%%%%%%%%%%%%%%%%

The expression (\ref{fs_res}) allows to analyze the main features of the critical temperature behavior. First, in contrast with the case of ${\rm S/F_1/F_2}$ spin valve\cite{Fominov_SFF} the dependence $T_c(\theta)$ is always symmetric, i.e. $T_c(\pi-\theta)=T_c(\theta)$. Second, $f_s\left(\sin^2\theta\right)$ is a monotonically decreasing function and, thus, the minimum of the critical temperature corresponds to $\theta=\pi/2$. Note that these two features were clearly observed in recent experiments with spin valves containing the half-metallic ${\rm CrO_2}$ layer.\cite{Aarts}

\begin{figure}[t!]
\includegraphics[width=0.38\textwidth]{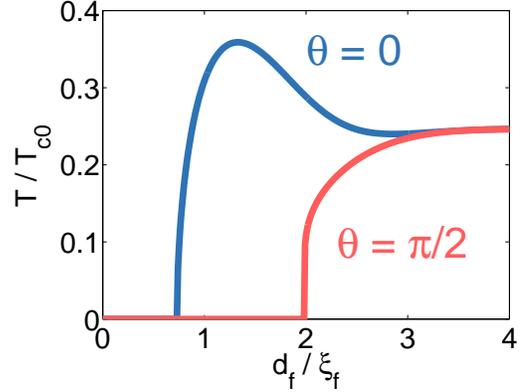}
\caption{(Color online) The dependencies of the critical temperature on the thickness of the ferromagnetic layer for $\theta=0$ (blue curve) and $\theta=\pi/2$ (red curve). The system parameters are the same as in Fig.~\ref{Fig_Tcphi}.} \label{Fig_Tcdf}
\end{figure}

The typical dependencies $T_c(\theta)$ are shown in Fig.~\ref{Fig_Tcphi}. One sees that S/F/HM system with $d_f\sim \xi_f$ reveal giant triplet spin-valve effect originating due to the LRTC: for the chosen parameters the increase of $\theta$ results in the damping of $T_c$ from $0.4T_{c0}$ to zero. In Fig.~\ref{Fig_Tcdf} we also plot the dependencies of $T_c$ on the F-layer thickness $d_f$ for $\theta=0$ and $\theta=\pi/2$. For $d_f\ll \xi_f$ the superconductivity is fully suppressed by the proximity with half-metal. When increasing $d_f$ above a certain threshold, which depends on $\theta$, the critical temperature becomes non-zero and grows rapidly. Interestingly, the difference in the thresholds for $\theta=0$ and $\theta=\pi/2$ makes it possible to simultaneously reach the absolute maximum of $T_c$ for $\theta=0$ and have $T_c(\pi/2)=0$.

%%%%%%%%%%%%%%%%%%%%%%%%%%%%%%%%%%%%%%%%%%%%
%%%%%%%%%%%%%%%%%%%%%%%%%%%%%%%%%%%%%%%%%%%%

\section{Anomalous Josephson effect in S/F/HM/F/S systems}\label{Sec_SFHMFS}

%%%%%%%%%%%%%%%%%%%%%%%%%%%%%%%%%%%%%%%%%%%%
\begin{figure}[b!]
\includegraphics[width=0.35\textwidth]{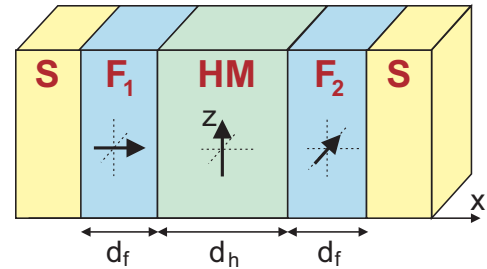}
\caption{(Color online) The sketch of the ${\rm S/F_1/HM/F_2/S}$ Josephson junction. The exchange field vectors in the ferromagnets and half-metal are perpendicular to each other.} \label{Fig_SFHMFS}
\end{figure}
%%%%%%%%%%%%%%%%%%%%%%%%%%%%%%%%%%%%%%%%%%%%

In this section we analyze the behavior of the Josephson current through the ${\rm F_1/HM/F_2}$-trilayer (see Fig.~\ref{Fig_SFHMFS}). In particular, we show that the peculiar mixing of different spin-triplet $\hat{f}$ components at F/HM interfaces results in the $\varphi_0$-junction formation provided the magnetic moments in the F-layers and the spin quantization axis in the HM-layer are the non-coplanar vectors. Previously this effect was noted within the circuit-theory for the junctions consisting of two ferromagnetic superconductors separated by the half-metal.\cite{Nazarov} The model of the ferromagnetic superconductors\cite{Nazarov} allows to simplify the calculations but can hardly be applied for the real systems in which the ferromagnetic order strongly suppresses superconductivity. Here we consider a more realistic situation when the regions with superconducting and ferromagnetic orders are separated in space.

To make the physical origin of the effects under consideration more transparent we restrict ourselves to the simplest case when the exchange field in the ${\rm F_1}$-layer is directed along the $x$-axis while the magnetic moment of the ${\rm F_2}$-layer has only $y$-component. For simplicity we consider equal magnitude $h$ of the exchange field in the ${\rm F_1}$ and ${\rm F_1}$-layers. Similar to the previous sections we assume that in the half-metal the spins are quantized in the $z$-direction. We choose the $x$-axis perpendicular to the interfaces so that the ${\rm F_1}$, ${\rm HM}$ and ${\rm F_2}$-layers occupies the regions $-(d_f+d_h/2)<x<-d_h/2$, $-d_h/2<x<d_h/2$ and $d_h/2<x<d_f+d_h/2$ respectively.

Before proceeding with the calculation of the Josephson current let us briefly point out the main difference of the described system from the ${\rm S_1/F_1/F'/F_2/S_2}$ junctions with the same magnetic configuration extensively studied before. In the latter system if the magnetic moments in the three ferromagnets are perpendicular to each other and the thickness of the central layer strongly exceeds the coherence length $\sqrt{D_2/h_2}$ the Josephson current is negligibly small. Indeed, the ${\rm F_1}$ layer produces the component $f_{tx}$ of the anomalous Green function which becomes long-range in the ${\rm F'}$ layer and reaches the ferromagnet ${\rm F_2}$. However, this component do not contribute to the Josephson current since the magnetic moment in the ${\rm F_2}$ directed along the $y$-axis cannot convert this $f_{tx}$ component into the singlet $f_s$ one.

The situation becomes completely different if instead of the ferromagnet ${\rm F'}$ one has half-metallic layer. In this case at the ${\rm F_1/HM}$ interface both $f_{tx}$ and $f_{ty}$ components are produced. As a result, in the ${\rm F_2}$-layer the $f_{ty}$ component can be effectively converted into the $f_s$ one and, thus, produce the non-vanishing Josephson current.

The current-phase relation $j(\varphi)$ of the Josephson junction is defined by the sum
\begin{equation}\label{Curr_def}
j\left(\varphi\right)=8\pi e D N T \sum\limits_{\omega_n>0}\Im\left(f_s^*\partial_x f_s-{\bf f}_t^*\partial_x {\bf f}_t\right),
\end{equation}
where $e$ is the charge of electron, $N$ is the electronic density of states, $D$ is the diffusion coefficient. Since the current does not depend on the position across the junction the anomalous Green function can be taken in arbitrary point. Practically it is convenient to choose this point, e.g., at the ${\rm S/F_1}$ interface (at $x=-d_f-d_h/2$) where only the singlet component $f_s$ is non-zero.

To simplify the further calculations we assume that the normal conductivity of the superconducting electrodes strongly exceed the ones in the ferromagnets, so that at both S/F interfaces the rigid boundary conditions are fulfilled: $f_s=(\Delta/\omega_n) e^{\pm i\varphi/2}$ (the signs $+$ and $-$ in the phase factor correspond to the right and left S-layers respectively) while ${\bf f}_t=0$. Also we assume that the thickness of the HM-layer is much less than the coherence length $\sqrt{D_h/4\pi T_{c0}}$ which allows to neglect the spatial variations of the function $f_{\uparrow\uparrow}$ across this layer. Solving the Usadel equation with the boundary conditions discussed above we obtain the analytical expression for the derivative $\partial_x f_s$ at $x=-d_f-d_h/2$ (see Appendix~\ref{App_CPR}) and the resulting current-phase relation:
\begin{equation}\label{CPR_res}
j\left(\varphi\right)=4\pi e D_f N T \sin\left(\varphi+\frac{\pi}{2}\right)\sum\limits_{\omega_n>0} \frac{\Delta^2}{\omega_n^2}J_n,
\end{equation}
where
\begin{equation}\label{CPR_res}
J_n=\frac{1}{q_f\coth(q_fd_f)+ \Re\left[q\coth(qd_f)\right]}\Im^2\left[\frac{q}{\sinh(qd_f)}\right]
\end{equation}
and the definition of $q$ and $q_f$ are the same as in Sec.~\ref{Sec_SFHM}.

From Eq.~(\ref{CPR_res}) one can see that the non-coplanarity of the magnetic moments in the magnetic layers results in the appearance of the spontaneous Josephson phase difference (so-called $\varphi_0$-junction). In contrast with the ordinary S/F/S systems where the formation of $\varphi_0$-junction requires strong spin-orbit, coupling\cite{Buzdin_Phi,Yokoyama,Mironov_BdG,Bergeret_Phi0} here the spontaneous phase arises due to spin polarization in the HM-layer. Note that in case of arbitrary mutual orientation of the magnetic moments in the F-layers one can expect the spontaneous Josephson phase $\varphi_0$ to be equal to the angle between the projections of these magnetic moments to the $xy$-plane (see Ref.~\onlinecite{Nazarov}).

\section{S/F/HM spin valve of the atomic thickness}\label{Sec_Atomic}

%%%%%%%%%%%%%%%%%%%%%%%%%%%%%%%%%%%%%%%%%%%%
\begin{figure}[t!]
\includegraphics[width=0.15\textwidth]{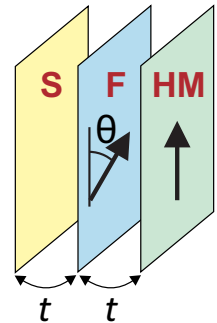}
\caption{(Color online) S/F/HM spin valve of the atomic thickness. The angle between the exchange field in the ferromagnet and the spin quantization axis in the half-metal is denoted as $\theta$. The coupling between the layers is described by the transfer integral $t$.} \label{Fig_SFHM_atomic}
\end{figure}
%%%%%%%%%%%%%%%%%%%%%%%%%%%%%%%%%%%%%%%%%%%%

In this section we study the spin-valve effect in the S/F/HM trilayers of atomic thickness. Experimentally such kind of systems can be realized, e.g., on the basis of the ${\rm RuSr_2GdCu_2O_8}$ and ${\rm La_{0.7}Ca_{0.3}MnO_3}$ compounds. The first one is the ferromagnetic superconductor with the alternating S-layers of ${\rm CuO_2}$ and the magnetically ordered ${\rm RuO_2}$ layers. The second compound is a half-metal\cite{CoeyAdv} which is recently shown to have a strong influence on the properties of the superconducting systems.\cite{Villegas}

Here we use the microscopical Gor'kov formalism to calculate how the critical temperature of the S/F/HM trilayer depends on the angle $\theta$ between the exchange field ${\bf h}=h\cos\theta\hat{\bf z}+h\sin\theta\hat{\bf x}$ in the ferromagnet and the spin quantization axis $\hat{\bf z}$ in the HM-layer (see Fig.~\ref{Fig_SFHM_atomic}). We assume that the neighboring layers are coupled by the electron tunneling describing by the transfer integral $t$. Previously this approach was successfully applied for the description of the spin-valve effect in F/S/F structures (see [\onlinecite{TollisAtomic,DaumensAtomic,Andreev,Prokic}] and references therein). The main advantage of the atomic layers model is the possibility to obtain the exact solutions for the Green functions and analyze their properties. Below we obtain such solutions for the case of the S/F/HM structures and demonstrate that they reproduce all main features of the spin-valve effect described within the phenomenological Usadel model in Sec.~\ref{Sec_SFHM}.

Let us denote the two-component electronic operators in the S, F and HM layers as $\varphi$, $\psi$ and $\eta$ respectively. For simplicity we assume that the quasiparticle motion in the plane of the S and F-layers is described by the same energy spectrum $\xi({\bf p})$. At the same time, in the half-metal the energy is strongly spin dependant: we assume that for the spin-up quasiparticles $\xi_\uparrow=\xi({\bf p})$ while for the spin-down ones $\xi_\downarrow=+\infty$, which implies $\eta_\downarrow=0$.

The Hamiltonian of the system under consideration has the form\cite{TollisAtomic}
\begin{equation}\label{Ham}
\hat{H}=\hat{H}_0+\hat{H}_{S}+\hat{H}_t,
\end{equation}
where
\begin{equation}\label{H0}
\hat{H}_0=\sum\limits_{{\bf p};\alpha=1,2}\left[\xi({\bf p})\varphi_\alpha^+\varphi_\alpha + \psi_\alpha^+\hat{C}(\theta)\psi_\alpha+ \eta_\alpha^+\hat{P}\eta_\alpha\right],
\end{equation}
\begin{equation}\label{HS}
\hat{H}_S=\sum\limits_{{\bf p}}\left(\Delta^*\varphi_{{\bf p},2}\varphi_{-{\bf p},1} + \Delta\varphi_{{\bf p},1}^+\varphi_{-{\bf p},2}^+\right),
\end{equation}
\begin{equation}\label{Ht}
\hat{H}_t=\sum\limits_{{\bf p};\alpha=1,2}t\left(\varphi_\alpha^+\psi_\alpha+ \psi_\alpha^+\varphi_\alpha +\psi_\alpha^+\eta_\alpha+ \eta_\alpha^+\psi_\alpha\right).
\end{equation}
In Eq.~(\ref{H0}) we introduces two matrices describing the effect of Zeeman coupling in the F-layer and the spin polarization in the half-metal:\cite{TollisAtomic}
\begin{equation}\label{HS}
\hat{C}(\theta)=\left(
                  \begin{array}{cc}
                    \xi-h\cos\theta & -h\sin\theta \\
                    -h\sin\theta & \xi+h\cos\theta \\
                  \end{array}
                \right),
~~~\hat{P}=\left(
                  \begin{array}{cc}
                    \xi & 0 \\
                    0 & +\infty \\
                  \end{array}
                \right).
\end{equation}
Note that in contrast with the model of Ref.~\onlinecite{Nazarov} in our system the superconducting and ferromagnetic regions are separated in space which makes it possible to consider the exchange field $h$ of arbitrary magnitude.

The critical temperature $T_c$ of the S-layer is defined by the linear expansion of the anomalous Green function $F_{\alpha,\beta}^+=\left<T_\tau\left(\varphi_\alpha^+,\varphi_\beta^+\right)\right>$ over the gap potential $\left|\Delta\right|$, where $T_\tau$ denotes the time-ordered product for the imaginary time $\tau$. Writing and solving the systems of Gor'kov equations for the S/F/HM system we find (the details of calculations are presented in Appendix~\ref{App_Gorkov}):
\begin{equation}\label{Gorkov_F_res}
\begin{array}{c}
{\ds \frac{\hat{F}^+}{\Delta^*}=\left\{(i\omega+\xi)\hat{1}-t^2\left[(i\omega+\hat{C})- t^2(i\omega+\hat{P})^{-1}\right]^{-1}\right\}^{-1}}\\{}\\{\ds \times\hat{I}\left\{(i\omega-\xi)\hat{1}-t^2\left[(i\omega-\hat{C})- t^2(i\omega-\hat{P})^{-1}\right]^{-1}\right\}^{-1}}.
\end{array}
\end{equation}
To simplify the further calculations we assume the tunneling constant $t$ to be small and perform the power expansion of Eq.~(\ref{Gorkov_F_res}) over $t^2$. To obtain the non-trivial dependence $T_c(\theta)$ we should keep the terms up to $t^6$. Also it is convenient to represent the self-consistency equation\cite{DaumensAtomic} in the form
\begin{equation}\label{Self_cons}
T_c(\theta)=T_c(0)-2T_{c0}^2\sum\limits_{\omega_n>0}\int\limits_{-\infty}^{+\infty}{\rm Re}\frac{F^+_{12}(\theta)-F^+_{12}(0)}{\Delta^*}d\xi,
\end{equation}
where $T_c(0)=T_{c0}\left[1-O(t^2)\right]$ is the critical temperature at $\theta=0$ [for $t\ll T_{c0}$ one has $\left|T_c(\theta)-T_{c0}\right|\ll T_{c0}$] and the sum is taken over the discrete set of positive Matsubara frequencies $\omega_n=\pi T_{c0}(2n+1)$.

Substituting the expression for $F^+_{12}$ into Eq.~(\ref{Self_cons}) we found:
\begin{equation}\label{DF_res}
T_c(\theta)=T_c(0)+\sum\limits_{\omega_n>0}\int\limits_{-\infty}^{+\infty}\frac{4T_{c0}^2t^6h^2\sin^2\theta d\xi}{(i\omega-\xi)(i\omega+\xi)^3\left[(i\omega+\xi)^2-h^2\right]^3},
\end{equation}
Taking the integral over $\xi$ we get:
\begin{equation}\label{GTc_res}
T_c(\theta)=T_c(0)-\sum\limits_{\omega_n>0}\frac{\pi T_{c0}^2t^6 h^2\sin^2\theta}{\omega^3(4\omega^2+h^2)^3},
\end{equation}

First, one can clearly see that the deviation of the critical temperature from $T_c(0)$ is proportional to $\sin^2\theta$ and, thus, $T_c(\pi-\theta)=T_c(\theta)$ in the full accordance with the conclusion of Sec.~\ref{Sec_SFHM}.

Second, Eq.~(\ref{GTc_res}) shows that the magnitude of the spin-valve effect which can be characterized by the value $\delta T_c=T_c(0)-T_c(\theta)$ has non-monotonic dependence on $h$ with the maximum $\delta T_c^{max}\propto t^6/T_{c0}^5$ at $h\sim T_{c0}$. Indeed, for $h\ll T_{c0}$ the exchange field weakly affect the system properties, and
\begin{equation}\label{GTc_res_h_small}
\delta T_c=\frac{511\zeta(9) t^6 h^2}{2^{15}\pi^8T_{c0}^7}\propto \frac{t^6 }{T_{c0}^5}\left(\frac{h}{T_{c0}}\right)^2.
\end{equation}
In the opposite limit when $h\gg T_{c0}$ the strong Zeeman splitting of the energy bands inside the F-layer effectively damps the tunneling constant between the layers, and as a result the spin-valve effect is also weak:
\begin{equation}\label{GTc_res_h_big}
\delta T_c=\frac{7\zeta(3)t^6}{8\pi^2T_{c0}h^4}\propto \frac{t^6}{T_{c0}^5}\left(\frac{T_{c0}}{h}\right)^4.
\end{equation}
In terms of the previously discussed Usadel theory the latter relation simply reflects the fact that for $h\gg T_{c0}$ the coherence length $\xi_f=\sqrt{D_f/h}$ in the F-layer becomes much smaller than its thickness, and superconducting correlations do not reach half-metal.

\section{Conclusion}\label{Sec_Conc}

To sum up, we proposed the phenomenological Usadel theory of the superconducting proximity effect in multilayered systems with a half-metallic layer. It is shown that the boundary between ferromagnet and half-metal serves as a source of additional triplet component of the anomalous Green function which is perpendicular to both exchange field ${\bf h}$ in the F-layer and the spin quantization axis in half-metal $\hat{\bf z}$. For the S/F/HM trilayes we analyzed the dependence of the critical temperature $T_c$ on the angle $\theta$ between ${\bf h}$ and $\hat{\bf z}$ and found that the discovered triplet component strongly enhances the spin valve effect compared to the traditional S/F/F structures: increasing the angle $\theta$ one can damp $T_c$ from the value comparable to the critical temperature of the isolated superconductor down to zero. Note that the described giant damping of $T_c$ appears {\it only} due to the long-range triplet correlations (LRTC) since the short range ones do not penetrate the HM-layer and, thus, are not sensitive to $\theta$. In addition, we showed that the full spin polarization in the HM-layer requires the symmetry $T_c(\pi-\theta)=T_c(\theta)$, which was clearly observed in recent experiments with the ${\rm MoGe/Cu/Ni/CrO_2}$ spin valves.\cite{Aarts}

To verify our main conclusions about the peculiarities of the spin-valve effect in the S/F/HM structures we considered the case when the layers of such system have the atomic thickness. For this case we obtained the exact analytical solution of the Gor'kov equations and calculated the dependencies $T_c(\theta)$. We found that if the tunneling rate between the layers is small the deviation of $T_c(\theta)$ from the critical temperature at $\theta=0$ is proportional to $\sin^2\theta$ which reproduces the symmetry relation found within the Usadel formalism.

Also we demonstrated that the new ``perpendicular'' triplet component of the anomalous Green function dramatically modifies the current-phase relation of the S/F/HM/F/S Josephson junctions provided the exchange field vectors in the F-layers and the spin quantization axis $\hat{\bf z}$ in half-metal are non-coplanar. First, we found that such systems support the $\varphi_0$-junction formation. This result is non-trivial since in the usual S/F/S and S/N/S structures the appearance of the spontaneous Josephson phase difference requires strong spin-orbit coupling. \cite{Buzdin_Phi, Yokoyama, Mironov_BdG, Bergeret_Phi0} In contrast, here such phase emerges only due to the spin selectivity of the half-metal. Second, the critical current of the S/F/HM/F/S structures does not vanish when the exchange field vectors and $\hat{\bf z}$ are perpendicular to each other. This result directly originates from the presence of the additional Green function component. It is exactly this component, which makes it possible not only to generate the long-range triplet correlations near the S-lead but also to convert them back into the singlet ones near the opposite lead. Note that previously these effects were discussed for the system of two ferromagnetic superconductors and the half-metal separated with the tunnel barriers.\cite{Nazarov} However the model of ferromagnetic superconductor is valid only for extremely weak exchange field values which make it inapplicable to the real heterostructures.

Finally, we would like to mention that the appearance of the additional triplet component of the anomalous Green function should strongly influence the local density of states (LDOS) and the electromagnetic response of all considered heterostructures. Indeed, spin-triplet correlations make positive contribution into the LDOS\cite{Buzdin_DOS,Kontos,Cottet_DOS,Nazarov} and, as a result, LDOS in the superconducting state can even exceed the one above $T_c$. Obviously, the new triplet component generated by the half-metal should provide more favorable conditions for the investigation of this unusual phenomenon as well as for the observation of the related Fulde-Ferrell-Larkin-Ovchinnikov instabilities.\cite{Mironov_FFLO}

\section*{ACKNOWLEDGMENTS}

The authors thank A. S. Mel'nikov for useful discussions. This work was supported by the French ANR ``MASH," NanoSC COST Action MP1201, and the Russian Presidential foundation (Grant SP-6340.2013.5).

\appendix

\section{Effective boundary conditions at F/HM interface}\label{App_BoundCond}

Let us derive the effective boundary condition for the $f_{tx}$ component of the anomalous Green function at the F/HM interface of the S/F/HM heterostructure. From Eq.~(\ref{Main_System}) using the condition $\partial_xf_{ty}=0$ at $x=-d_S$ and $x=d_F+d_H$ one finds the solution for the $f_{ty}$ in the S and F layers as well as the solution for $f_{\uparrow\uparrow}$ in the HM-layer:
\begin{equation}\label{fy_GenSol}
\begin{array}{l}{\ds {\rm S:}~~~~~~ f_{ty}=A\cosh\left[q_s\left(x+d_s\right)\right],}\\{\ds {\rm F:}~~~~~~ f_{ty}=B\cosh\left[q_f\left(x-d_f\right)\right]+C\sinh\left[q_f\left(x-d_f\right)\right],}\\{\ds {\rm HM:}~~~f_{\uparrow\uparrow}=H\cosh\left[q_h\left(x-d_f-d_h\right)\right],}
\end{array}
\end{equation}
where $q_j=\sqrt{2\omega_n/D_j}$ (the index $j=s,f,h$ corresponds to the S, F and HM layer respectively), $D_j$ is the diffusion constant in the $j$-th layer, and $A$, $B$, $C$ and $H$ are the integration constants. Introducing the parameter $\nu_{sf}=\left(\sigma_s/\sigma_f\right)\sqrt{D_f/D_s}$ and $\nu_{hf}=\left(\sigma_h/\sigma_f\right)\sqrt{D_f/D_h}$ and using the boundary conditions at $x=0$ and $x=d$ we obtain the systems of equations:
\begin{equation}
\begin{array}{c}{\ds
B\cosh(q_fd_f)-C\sinh(q_fd_f)=A\cosh\left(q_sd_s\right),}\\{\ds
-B\sinh(q_fd_f)+C\cosh(q_fd_f)=\nu_{sf} A\sinh\left(q_sd_s\right),}\\{\ds
f_{tx}(d_f)+iB=0,}\\{\ds
-f_{tx}(d_f)+iB=H\cosh\left(q_hd_h\right),}\\{\ds
-\partial_xf_{tx}(d_f)+iq_fC=-q_f\nu_{hf}H\sinh\left(q_hd_h\right).}
\end{array}
\end{equation}
Excluding the constants $A$, $B$, $C$ and $H$ from this system we obtain the effective boundary condition (\ref{fx_bound}) for the component $f_{tx}$.

\section{Calculation of the anomalous Green function in the S/F/HM spin valve}\label{App_Calc}

As follows from Eq.~(\ref{Main_System}) if the gap potential $\Delta$ is chosen to be real then the components $f_s$ and $f_{ty}$ of the anomalous Green function are also real while the components $f_{tz}$ and $f_{tx}$ are imaginary. Then it is convenient to introduce the complex function
\begin{equation}\label{F_def}
F(x)=f_s+f_{tz}\cos\theta+f_{tx}\sin\theta
\end{equation}
and the real function
\begin{equation}\label{R_def}
R(x)=i\left(f_{tz}\sin\theta-f_{tx}\cos\theta\right),
\end{equation}
so that
\begin{equation}\label{f_from_FR}
\begin{array}{l}
{\ds f_s=\Re (F),}\\{\ds f_{tz}=i\Im(F)\cos\theta-iR\sin\theta ,}\\{\ds f_{tx}=i\Im(F)\sin\theta+iR\cos\theta.}
\end{array}
\end{equation}
The introduced functions satisfy the systems of equations
\begin{equation}\label{FR_system}
\begin{array}{l}{\ds \frac{D}{2}\partial_x^2 F=\left(\omega_n +ih\right)F-\Delta,}\\{}\\{\ds \frac{D}{2}\partial_x^2 R=\omega_n R.}
\end{array}
\end{equation}
The boundary conditions for $F$ and $R$ straightly follow from the corresponding conditions for the $\hat{f}$ components.

Since we assumed that $d_s\ll\xi_{s0}$ the solution of Eq.~(\ref{FR_system}) inside the superconductor satisfying the boundary conditions at $x=-d_s$ can be represented in the form
\begin{equation}\label{FR_S_solution}
\begin{array}{l}{\ds F=\frac{\Delta}{\omega_n}+F_S\frac{q_s^2(x+d_s)^2}{2},}\\{}\\{\ds R=R_S\frac{q_s^2(x+d_s)^2}{2}.}
\end{array}
\end{equation}
The analogous solution in the F layer reads
\begin{equation}\label{FR_F_solution}
\begin{array}{l}{\ds F=F_1\cosh\left[q(x-d_f)\right]+F_2 \sinh\left[q(x-d_f)\right],}\\{}\\{\ds R=R_1\cosh\left[q_f(x-d_f)\right]+R_2 \sinh\left[q_f(x-d_f)\right],}
\end{array}
\end{equation}
where we have introduced the complex wave-vector $q=\sqrt{2\left(\omega_n+ih\right)/D_f}$.

To calculate the unknown amplitudes in the functions $F$ and $R$ let us first consider the boundary conditions at $x=d_f$. Since $f_s=f_{tz}=0$ at this interface one finds:
\begin{equation}\label{BC_FHM_sz}
\Re(F_1)=0,~~~ \Im(F_1)\cos\theta=R_1\sin\theta.
\end{equation}
These two equations can be rewritten as
\begin{equation}\label{BC_FHM_sz_c}
F_1=iR_1\tan\theta.
\end{equation}
The condition for the component $f_{tx}$ reads [see (\ref{fx_bound})]
\begin{equation}\label{BC_FHM_x}
\Im(qF_2)\sin\theta+q_fR_2\cos\theta=-q_f\Gamma\left[\Im(F_1)\sin\theta+R_1\cos\theta\right].
\end{equation}

The boundary conditions at $x=0$ give the rest four equations:
\begin{equation}\label{BC_SF}
\begin{array}{l}{\ds \frac{\Delta}{\omega_n}+F_S=F_1\cosh(qd_f)-F_2 \sinh(qd_f),}\\{}\\{\ds pF_S=-q F_1\sinh(qd_f)+qF_2 \cosh(qd_f),}\\{}\\{\ds R_S=R_1\cosh(q_fd_f)-R_2 \sinh(q_fd_f),}\\{}\\{\ds pR_S=-q_fR_1\sinh(q_fd_f)+q_fR_2 \cosh(q_fd_f),}
\end{array}
\end{equation}
where
\begin{equation}\label{mu_approx}
p=\mu_sq_f\approx\frac{\sigma_s}{\sigma_f}\frac{2d_s}{D_s}\omega_n.
\end{equation}

To solve the system of equations (\ref{BC_FHM_sz_c})-(\ref{BC_SF}) it is convenient to exclude $F_S$ and $R_S$ from (\ref{BC_SF}) and then express $R_2$ and $F_2$ in terms of $R_1$ using (\ref{BC_FHM_sz_c}). The result is [see Eq.~(\ref{Gamma_def})]
\begin{equation}\label{R2F2}
\begin{array}{l}{\ds R_2=\left(\Gamma-2\mu_h\right)R_1,}\\{}\\{\ds F_2=\frac{iR_1\tan\theta\left[p\cosh(qd_f)+q\sinh(qd_f)\right]-p\Delta/\omega_n}{q\cosh(qd_f)+p\sinh(qd_f)}.}
\end{array}
\end{equation}
Finally, substituting (\ref{R2F2}) into (\ref{BC_FHM_x}) we obtain:
\begin{equation}\label{R1}
R_1=\frac{\Delta}{\omega_n}\frac{\mu_sW\sin\theta\cos\theta}{Q\sin^2\theta +\Gamma+\left(\Gamma-2\mu_h\right)\cos^2\theta},
\end{equation}
where
\begin{equation}\label{W_def}
W=\Im\left\{\frac{q}{q\cosh(qd_f)+p\sinh(qd_f)}\right\},
\end{equation}
\begin{equation}\label{Q_def}
Q=\Re \left\{\frac{q}{q_f}\frac{p+q\tanh(qd_f)}{q+p\tanh(qd_f)}\right\}.
\end{equation}
The obtained explicit expression for $R_1$ enables straightforward calculation of all other amplitudes in the anomalous Green function. In particular,
\begin{equation}\label{Fs_calc}
F_S=\frac{q\left[iR_1\tan\theta-(\Delta/\omega_n)\cosh(qd_f)\right]}{q\cosh(qd_f)+p\sinh(qd_f)}.
\end{equation}

\section{Calculation of the current-phase relation for the S/F/HM/F/S junction}\label{App_CPR}

Let us denote the coordinates of the left and right boundaries of the HM-layer as $x_L=-d_h/2$ and $x_R=d_h/2$. It is convenient to represent the solution of the Usadel equation in the ${\rm F_1}$-layer in the form
\begin{equation}\label{JJ_F1_Gen_Sol}
\begin{array}{l}{\ds f_s+ f_{tx}=A_1^+\sinh\left[q\left(x-x_L\right)\right]+ B_1\cosh\left[q\left(x-x_L\right)\right],}\\{}\\{\ds f_s- f_{tx}=A_1^-\sinh\left[q^*\left(x-x_L\right)\right]- B_1\cosh\left[q^*\left(x-x_L\right)\right],}\\{}\\{\ds f_{ty}= iB_1\frac{\sinh\left[q_f\left(x-x_L+d_f\right)\right]}{\sinh(q_fd_f)},}\\{}\\{\ds f_{tz}=0,}
\end{array}
\end{equation}
where $q_f=\sqrt{2\omega_n/D_f}$ and $q=\sqrt{2(\omega_n+ ih)/D_f}$ (we assume that the diffusion constants and normal conductivities in the F-layers are equal to each other). In Eq.~(\ref{JJ_F1_Gen_Sol}) we took into account that $f_s=f_{\downarrow\downarrow}=0$ at $x=x_L$ and $f_{ty}=0$ at $x=x_L-d_f$. Analogously, the solution in the ${\rm F_2}$-layer can be written as
\begin{equation}\label{JJ_F2_Gen_Sol}
\begin{array}{l}{\ds f_s+ f_{ty}=A_2^+\sinh\left[q\left(x-x_R\right)\right]+B_2\cosh\left[q\left(x-x_R\right)\right],}\\{}\\{\ds f_s- f_{ty}=A_2^-\sinh\left[q^*\left(x-x_R\right)\right]-B_2\cosh\left[q^*\left(x-x_R\right)\right],}\\{}\\{\ds f_{tx}= iB_2\frac{\sinh\left[q_f\left(x-x_R-d_f\right)\right]}{\sinh(q_fd_f)},}\\{}\\{\ds f_{tz}=0.}
\end{array}
\end{equation}

Inside the HM-layer the solution for the only non-zero component $f_{\uparrow\uparrow}=-f_{tx}+if_{ty}$ has the form
\begin{equation}\label{JJ_HM_Gen_Sol}
f_{\uparrow\uparrow}=P_1\sinh(q_hx)+P_2\cosh(q_hx),
\end{equation}
with $q_h=\sqrt{2\omega_n/D_h}$. Taking derivation of Eq.~(\ref{JJ_HM_Gen_Sol}) and excluding the constants $P_1$ and $P_2$ we obtain two equations which connect the values of $f_{\uparrow\uparrow}$ and $\partial_xf_{\uparrow\uparrow}$ on the left and right sides of the half-metal (the corresponding values are indicated by the upper indexes $L$ and $R$):
\begin{equation}\label{HM_LR}
\begin{array}{l}{\ds
q_hf_{\uparrow\uparrow}^R=\partial_xf_{\uparrow\uparrow}^L\sinh(q_hd_h)+q_hf_{\uparrow\uparrow}^L\cosh(q_hd_h),}\\{}\\{\ds \partial_xf_{\uparrow\uparrow}^R=\partial_xf_{\uparrow\uparrow}^L\cosh(q_hd_h)+q_hf_{\uparrow\uparrow}^L\sinh(q_hd_h).}
\end{array}
\end{equation}
Further for simplicity we will assume that $d_h\ll\sqrt{D_h/4\pi T_{c0}}$. Then the system (\ref{HM_LR}) transforms into $f_{\uparrow\uparrow}^R=f_{\uparrow\uparrow}^L$ and $\partial_xf_{\uparrow\uparrow}^R=\partial_xf_{\uparrow\uparrow}^L$. Taking this into account and substituting Eqs.~(\ref{JJ_F1_Gen_Sol})-(\ref{JJ_F2_Gen_Sol}) into the boundary conditions at the S/F and F/HM interfaces we obtain:
\begin{equation}\label{JJ_E1}
-A_1^+\sinh(qd_f)+ B_1\cosh(qd_f) = \frac{\Delta }{\omega_n}e^{-i\varphi/2},
\end{equation}
\begin{equation}
-A_1^-\sinh(q^*d_f)- B_1\cosh(q^*d_f) = \frac{\Delta }{\omega_n}e^{-i\varphi/2},
\end{equation}
\begin{equation}
A_2^+\sinh(qd_f)+B_2\cosh(qd_f)=\frac{\Delta }{\omega_n}e^{i\varphi/2},
\end{equation}
\begin{equation}
A_2^-\sinh(q^*d_f)-B_2\cosh(q^*d_f)=\frac{\Delta }{\omega_n}e^{i\varphi/2},
\end{equation}
\begin{equation}
-B_1=iB_2,
\end{equation}
\begin{equation}
-B_1=iB_2
\end{equation}
\begin{equation}\label{JJ_E2}
\begin{array}{c}
{\ds -qA_1^++q^*A_1^--2B_1q_f\coth(q_fd_f)=~~~~~}\\{}\\{\ds ~~~~~~~-2iB_2q_f\coth(q_fd_f)+iqA_2^+-iq^*A_2^-.}
\end{array}
\end{equation}

To calculate the Josephson current through the junction we need to find only the combination $\Im\left(f_s^*\partial_x f_s\right)$ at $x=-d_f-d_h/2$. Solving the system of equations (\ref{JJ_E1})-(\ref{JJ_E2}) we find:
\begin{equation}\label{B1_res}
B_1=\frac{1}{2}\frac{ie^{-i\varphi/2}+e^{i\varphi/2}}{q_f\coth(q_fd_f)+ \Re\left[q\coth(qd_f)\right]}\Im\left[\frac{q}{\sinh(qd_f)}\right]
\end{equation}
and, as a consequence,
\begin{equation}\label{Der_res}
\Im\left(f_s^*\partial_x f_s\right)=\frac{1}{2}\frac{\sin\left(\varphi+\pi/2\right)\Im^2\left[\frac{q}{\sinh(qd_f)}\right]}{q_f\coth(q_fd_f)+ \Re\left[q\coth(qd_f)\right]}
\end{equation}
Substituting this expression into Eq.~(\ref{Curr_def}) we obtain the desired current-phase relation.

\section{Calculation of the Green function for the S/F/HM system of atomic thickness}\label{App_Gorkov}

Let us introduce the following Green functions in the imaginary time representation:
\begin{equation}\label{Green_def}
\begin{array}{c}{\ds
G_{\alpha,\beta}=-\left<T_\tau\left(\varphi_\alpha,\varphi_\beta^+\right)\right>, F_{\alpha,\beta}^+=\left<T_\tau\left(\varphi_\alpha^+,\varphi_\beta^+\right)\right>,}\\{}\\
{\ds
E_{\alpha,\beta}^{\psi}=-\left<T_\tau\left(\psi_\alpha,\varphi_\beta^+\right)\right>, F_{\alpha,\beta}^{\psi+}=\left<T_\tau\left(\psi_\alpha^+,\varphi_\beta^+\right)\right>,}\\{}\\
{\ds
E_{\alpha,\beta}^{\eta}=-\left<T_\tau\left(\eta_\alpha,\varphi_\beta^+\right)\right>, F_{\alpha,\beta}^{\eta+}=\left<T_\tau\left(\eta_\alpha^+,\varphi_\beta^+\right)\right>,}
\end{array}
\end{equation}
Then performing the Fourier transform we obtain the following system of the matrix Gor'kov equations:
\begin{equation}\label{Gorkov1}
\left(i\omega-\xi\right)\hat{G}-t\hat{E}^\psi+\Delta\hat{I}\hat{F}^+=\hat{1},
\end{equation}
\begin{equation}\label{Gorkov2}
\left(i\omega+\xi\right)\hat{F}^++t\hat{F}^{\psi+}-\Delta^*\hat{I}\hat{G}=0,
\end{equation}
\begin{equation}\label{Gorkov3}
\left(i\omega-\hat{C}\right)\hat{E}^\psi-t\hat{G}-t\hat{E}^\eta=0,
\end{equation}
\begin{equation}\label{Gorkov4}
\left(i\omega+\hat{C}\right)\hat{F}^{\psi+}+t\hat{F}^++t\hat{F}^{\eta+}=0,
\end{equation}
\begin{equation}\label{Gorkov5}
\left(i\omega-\hat{P}\right)\hat{E}^\eta-t\hat{E}^\psi=0,
\end{equation}
\begin{equation}\label{Gorkov6}
\left(i\omega+\hat{P}\right)\hat{F}^{\eta+}+t\hat{F}^{\psi+}=0,
\end{equation}
where $\hat{I}=i\hat{\sigma}_y$, and $\hat{1}$ is the unit matrix in the spin space. Solving the system (\ref{Gorkov1})-(\ref{Gorkov6}) and considering only the linear term in the expansion of the function $\hat{F}^+$ over $\left|\Delta\right|$ we obtain the expression (\ref{Gorkov_F_res}).

\end{document}